\documentstyle[12pt,epsf]{article}

\advance\voffset by -1.5cm
\advance\hoffset by -2cm
\def\be{\begin{equation}}
\def\ee{\end{equation}}

\def\pmb#1{\setbox0=\hbox{#1}
 \kern-.025em\copy0\kern-\wd0
 \kern.05em\copy0\kern-\wd0
 \kern-.025em\raise.0433em\box0 }

\def\3{\ss}
\def\sq{\hbox{\rlap{$\sqcap$}$\sqcup$}}
\def\qed{\ifmmode\sq\else{\unskip\nobreak\hfil
\penalty50\hskip1em\null\nobreak\hfil\sq
\parfillskip=0pt\finalhyphendemerits=0\endgraf}\fi}

\def\bbbz {{\sf Z\!\!Z}}

\def\bbbr {{\rm I\!R}}

\def\RR{R$-$R }
\def\NSNS{NS$-$NS }

\newcommand{\ket}[1]{|#1\rangle}

\def\R{{\cal R}}

\def\ss{\bf S}

\def\F{{\cal F}}
\def\R{{\cal R}}

\def\drawbox#1#2{\hrule height#2pt 
        \hbox{\vrule width#2pt height#1pt \kern#1pt 
              \vrule width#2pt}
              \hrule height#2pt}

\def\Fund#1#2{\vcenter{\vbox{\drawbox{#1}{#2}}}}
\def\Asym#1#2{\vcenter{\vbox{\drawbox{#1}{#2}
              \kern-#2pt       
              \drawbox{#1}{#2}}}}

\def\funda{\Fund{6.5}{0.4}}

\def\symm{\funda\kern-0.4pt\funda}

\begin{document}

\thispagestyle{empty}
\def\thefootnote{\fnsymbol{footnote}}
\begin{flushright}
  hep-th/0001130\\
  CALT-68-2256 \\
  CITUSC/00-002 \\
  DAMTP-2000-4
 \end{flushright}
\vskip 0.5cm

\begin{center}\LARGE
{\bf On the Consistency of Orbifolds}
\end{center}
\vskip 0.7cm
\begin{center}
{\large  Oren Bergman\footnote{E-mail  address: 
{\tt bergman@theory.caltech.edu}}}

\vskip 0.5 cm
{\it California Institute of Technology,
Pasadena, CA 91125, USA}\\
and\\
{\it CIT/USC Center for Theoretical Physics, Univ. of Southern California,
Los Angeles CA}

\vskip 0.7 cm
{\large  Matthias R. Gaberdiel\footnote{E-mail  address: 
{\tt M.R.Gaberdiel@damtp.cam.ac.uk}}}

\vskip 0.5 cm
{\it Department of Applied Mathematics and Theoretical Physics \\
Centre for Mathematical Sciences \\
Wilberforce Road \\
Cambridge CB3 0WA, U.K.}
\end{center}

\vskip 0.5cm

\begin{center}
January 2000
\end{center}

\vskip 0.5cm

\begin{abstract}
Modular invariance is a necessary condition for the consistency of any
closed string theory. In particular, it imposes stringent constraints
on the spectrum of orbifold theories, and in principle determines their 
spectrum uniquely up to discrete torsion classes. In practice,
however, there are often ambiguities in the construction of orbifolds
that are a consequence of the fact that the action of the orbifold
elements on degenerate ground states is not unambiguous. We explain
that there exists an additional consistency condition, related to the
spectrum of D-branes in the theory, which eliminates these
ambiguities. For supersymmetric orbifolds this condition turns out to
be equivalent to the condition that supersymmetry is unbroken in the
twisted sectors, but for non-supersymmetric orbifolds it appears to be
a genuinely new consistency condition.
\end{abstract}

\vskip 0.8cm 
\begin{center}
\end{center}

\vfill
\setcounter{footnote}{0}
\def\thefootnote{\arabic{footnote}}
\newpage

\paragraph{1. Introduction.}

The orbifold construction is one of the most powerful techniques in
string theory that allows one to obtain interesting new theories with
reduced symmetries \cite{DHVW}. This procedure can be applied whenever
a string theory possesses a discrete symmetry group $\Gamma$. The
construction proceeds in two steps: first we restrict the space of
states to those that are invariant under $\Gamma$, and then we add
suitably projected twisted sectors, as needed by modular invariance.  
The initial projection is equivalent to adding sectors to the
partition function in which the boundary condition in the timelike 
world-sheet direction $t$ is twisted by generators of $\Gamma$. 
One of the generators of the modular group of the torus,
$S:\tau\mapsto -1/\tau$, exchanges the timelike direction $t$ with the
spacelike world-sheet direction $\sigma$. In order for the partition
function to be modular invariant, the theory must therefore also
contain sectors in which the boundary condition in $\sigma$ is twisted
by the generators of $\Gamma$; these are the twisted sectors. Finally,
invariance under the other generator of the modular group,
$T:\tau\mapsto\tau+1$, requires us to add sectors in which both
boundary conditions are twisted; in essence this amounts to projecting
the twisted sectors by $\Gamma$ as well. 

Whilst this algorithm determines the theory in principle, 
there are a number of ambiguities that are not fixed by modular
invariance of the one-loop vacuum amplitude alone. For example, if we
are dealing with a string theory in the RNS formalism, every sector of
the theory must be GSO-projected. In each sector that possesses
fermionic zero modes, and therefore a degenerate ground state, 
the GSO-projection acts as the chirality operator on the ground state. 
{\it A priori}, the definition of this operator is only fixed up to a
sign, and this sign choice affects the spectrum of the theory
significantly. On the other hand, the partition function is
insensitive to this choice, and therefore the condition of one-loop
vacuum modular invariance does not fix this ambiguity. 

It is generally believed that these ambiguities are resolved by the
conditions that come from non-vacuum and higher-loop amplitudes, and
that the remaining freedom is described in terms of discrete torsion
\cite{Vafa}. In practice, it is however difficult to determine the
consistent GSO-projections explicitly. In fact, the analysis of
\cite{Vafa} describes the freedom in modifying the action of the
orbifold (and GSO-projections) in the twisted sectors {\em relative}
to a certain solution which is assumed to be consistent, namely the
one in which all phases are taken to be $+1$. 
In particular, this requires that the above sign ambiguities have been
chosen so as to give a consistent solution, but it is {\it a priori}
not obvious how this should be done. 

As we shall explain in this paper, there exists a simple
{\em non-perturbative} consistency condition that fixes these
ambiguities, at least for a certain class of theories, uniquely.
This condition arises from a careful analysis of the D-brane spectrum
of these theories. The D-brane spectrum can be determined using the
boundary state formalism \cite{PolCai,CLNY,Cardy,Lew,BG1,GS}, in which
D-branes are described by coherent states of the closed string
sector. Once the various projection operators of the closed string
theory have been fixed, the D-brane spectrum is uniquely
determined. However, not every such spectrum is allowed, since there
exist transitions between different D-brane configurations. 
Specifically, when a D-brane in the bulk collides with a fixed-plane
of the orbifold it must be allowed to break into fractional
D-branes. In order for this to occur, however, both types of branes
must exist in the spectrum. As we shall see, this requirement fixes
the ambiguities in the GSO-projection of the twisted sectors
uniquely. 

For orbifold projections that preserve supersymmetry, these ambiguities
can also be fixed by requiring the twisted sectors to preserve the
same supersymmetry. This is the case, for example, for Type IIA or IIB
on $T^4/\bbbz_2$, where the $\bbbz_2$ generator acts by reflection of
the four compact coordinates. In these cases, the supersymmetry
considerations pick out precisely the theory that is also
non-perturbatively 
consistent. For orbifolds that break supersymmetry completely,
however, there does not appear to be an analogous perturbative
criterion that selects the non-perturbatively consistent theory.   

Our interest in this problem arose from recent work of Klebanov,
Nekrasov and Shatashvili \cite{KNS}, who considered Type IIB on
$T^6/\bbbz_4$, where the $\bbbz_4$ generator acts by the reflection of
the six compact directions. This orbifold breaks all the
supersymmetries of Type IIB, and is in fact equivalent to Type 0B on
$T^6/\bbbz_2$.  As we shall discuss in detail below, their choice for 
the GSO-projection in the twisted sectors does not satisfy the
above condition, and therefore does not define a (non-perturbatively) 
consistent theory. This resolves the question of which state is
charged under the $U(1)$ gauge field that comes from the twisted R-R 
sector, since the consistent theory does not have such a gauge field.

\paragraph{2. D-branes on $\bbbz_2$ orbifolds.} 

Let us begin by describing briefly the D-brane spectrum of a $\bbbz_2$
orbifold, where the $\bbbz_2$ generator acts by the inversion of $n$
coordinates. (More details can be found in \cite{GS}; see also
\cite{DM,Sen2,BG2,Sen6,BG3,DG,GabSen}.) 
For simplicity we shall consider the non-compact case
$\bbbr^n/\bbbz_2$, which only has one fixed plane (of spatial
dimension $9-n$) at the origin.\footnote{If $m$ of the $n$ directions
on which the $\bbbz_2$ acts are compact there will be $2^m$ fixed
planes.}  There exist three possible types of D-branes, which differ
in their boundary state components. {\bf Bulk} D-branes have
components only in the untwisted sectors,  
\be
\label{bulk}
\ket{Dp}_b = \Big(\ket{Bp}_{\mbox{\scriptsize\NSNS;U}} +
 \ket{Bp}_{\mbox{\scriptsize\RR;U}}\Big)\,, 
\ee
{\bf fractional} D-branes have components in all untwisted and twisted
sectors, 
\be
\label{fractional}
\ket{Dp}_f = {1\over2} \Big(\ket{Bp}_{\mbox{\scriptsize\NSNS;U}} 
+ \ket{Bp}_{\mbox{\scriptsize\RR;U}}
+ \ket{Bp}_{\mbox{\scriptsize\NSNS;T}} 
+ \ket{Bp}_{\mbox{\scriptsize\RR;T}} \Big)\,,
\ee
and {\bf truncated} D-branes only involve the untwisted NS-NS and
twisted R-R sectors, 
\be
\label{nonBPS}
\ket{Dp}_t = {1\over\sqrt{2}}
\Big(\ket{Bp}_{\mbox{\scriptsize\NSNS;U}} 
+ \ket{Bp}_{\mbox{\scriptsize\RR;T}} \Big)\,,
\ee
and can only exist for values of $p$ for which a fractional (and bulk) 
D-brane does not exist. The above spectrum, including the latter
restriction, follows from the usual condition that the cylinder
amplitude between any two D-branes must correspond to an open string
partition function. 

In a supersymmetric theory (\ref{bulk}) and (\ref{fractional}) are BPS
states, whereas (\ref{nonBPS}) are non-BPS, but nevertheless stable in
certain regimes of the moduli space \cite{Sen6,BG3}. To account for
the different orientations of the D-branes relative to the action of
the orbifold, we denote as in \cite{GS} the number of world-volume
directions parallel to the fixed-plane by $r$, and the number of
world-volume directions transverse to the fixed-plane by $s$. We shall
henceforth label the D$p$-branes (and boundary states) by $(r,s)$,
where $p=r+s$.  Fractional and truncated branes are either completely
localised at the fixed-plane (if $s=0$), or extend in $s$ directions
transverse to the fixed-plane and terminate at an $r$-dimensional
hyperplane in it; this follows from the fact that the boundary states
have components in the twisted sectors.\footnote{In the compact case
fractional and truncated branes would contain components in multiple
twisted sectors, and would therefore be suspended between different
fixed planes.}

Physical closed string states must be invariant under the
GSO-projection and the action of the orbifold group.  The spectrum of
{\em physical} D-branes, {\em i.e.} the allowed values of $r$ and $s$
for the three types of branes, is therefore determined by the action
of the GSO and orbifold operators on the boundary states in the four
different sectors. Since these boundary states are generically given
by  
\be
 \ket{B(r,s)} = e^{\sum(\pm{1\over n}
        \alpha_{-n}^i\widetilde{\alpha}_{-n}^i 
 \pm \psi_{-r}^i\widetilde{\psi}_{-r}^i)} \ket{B(r,s)}^{(0)}\;,
\ee
the crucial constraint comes from the action of the above operators on
the ground states of the different sectors. The left- and right-moving
GSO-operators are given as  
\be
 (-1)^f = \pm (i^{|\F|/2}) \prod_{\mu\in\F} (\sqrt{2} \psi_0^\mu)
 \qquad\qquad 
 (-1)^{\widetilde{f}} = \pm (i^{|\F|/2}) \prod_{\mu\in\F} 
(\sqrt{2} \widetilde{\psi}_0^\mu) \,,
\label{GSO-action}
\ee
where $\psi^\mu_0$ and $\widetilde{\psi}^\mu_0$ are the left- and
right-moving fermionic zero modes, respectively, which satisfy the
Clifford algebra 
\be
\{\psi^\mu_0,\psi^\nu_0\} = \eta^{\mu\nu} \qquad
\{\psi^\mu_0,\widetilde\psi^\nu_0\} = 0 \qquad
\{\widetilde\psi^\mu_0,\widetilde\psi^\nu_0\} = \eta^{\mu\nu}\,,
\ee
and $\F$ denotes the set of coordinates in which the given sector
has fermionic zero modes. The prefactor has been fixed (up to a sign)
so that both operators square to the identity (we shall only consider
the case where the number of elements in ${\cal F}$, $|{\cal F}|$, is
even). In the untwisted NS-NS sector, for which $|{\cal F}|=0$, 
it is conventional to take both signs to be $-$, so that the tachyonic
ground state is odd under both operators. For definiteness we adopt
this convention for all other sectors as well. Finally, the inversion
operator of the orbifold acts as    
\be
 g = \prod_{\mu\in\F\cap \R} (\sqrt{2}\psi_0^\mu)
     \prod_{\mu\in\F\cap \R} (\sqrt{2}\widetilde\psi_0^\mu) \,,
\label{orb-action}
\ee
where $\R$ denotes the set of coordinates on which the orbifold acts.
Again, the definition of $g$ in the twisted sectors is {\it a priori}
ambiguous up to a sign; for definiteness we have again fixed this to
be $+$ in all sectors. The resulting actions on the boundary
states\footnote{One must actually combine boundary states with
different spin structures to get GSO-eigenstates; the listed states
are the linear combination for which $(-1)^f$ has eigenvalue $+1$.}
are shown in table~1 (see \cite{GS} for details).
\begin{table}[htb]
\centering
\begin{tabular}{|l|l|l|l|}
\hline
 boundary state & $(-1)^f$ & $(-1)^{\widetilde{f}}$ & $g$ \\
\hline
 &&&\\[-10pt]
 $\ket{B(r,s)}_{\mbox{\scriptsize\NSNS;U}}$ & $+1$ & $+1$ & $+1$ \\[5pt]
 $\ket{B(r,s)}_{\mbox{\scriptsize\RR;U}}$ & $+1$ & $(-1)^{r+s+1}$ & 
   $(-1)^{s}$ \\[5pt]
 $\ket{B(r,s)}_{\mbox{\scriptsize\NSNS;T}}$ & $+1$ & $i^n(-1)^{s}$ & 
   $(-1)^{s}$ \\[5pt]
 $\ket{B(r,s)}_{\mbox{\scriptsize\RR;T}}$ & $+1$ & $i^n(-1)^{r+1}$ & 
  $+1$ \\[5pt]
\hline
\end{tabular}
\caption{\small
GSO and orbifold actions on the boundary states.}
\end{table}

The definition of the GSO and orbifold projections in the various sectors 
determines which of the boundary states are physical, and therefore
the D-brane spectrum of the theory. However, in order for this D-brane
spectrum to make sense, it must satisfy an additional consistency
condition:  as a bulk D$(r,s)$-brane brane approaches the fixed plane,
additional massless scalars usually appear in the world-volume gauge
theory; these parametrise the Coulomb branch, and describe the moduli
along which the bulk brane fractionates into two fractional branes. In
order for this to be possible, the theory must therefore also have a
fractional D$(r,s)$-brane. As we shall see below, this condition fixes
uniquely the ambiguity in defining the GSO-projection in the  twisted
sectors. 

We shall illustrate this condition by considering two examples.
The first is Type II on $\bbbr^4/\bbbz_2$, which is supersymmetric,
and the second is Type 0 on $\bbbr^6/\bbbz_2$ (or equivalently Type II
on $\bbbr^6/\bbbz_4$), in which supersymmetry is broken.

\paragraph{3. Supersymmetric example.}

Consider Type II strings on $\bbbr^{1,5}\times\bbbr^4/\bbbz_2$, 
where the generator $g$ of $\bbbz_2$ reflects the coordinates
$x^5,x^6,x^7,x^8$. This breaks $SO(1,9)$ to $SO(1,5)\times SO(4)_R$
($SO(4)_S\times SO(4)_R$ in light-cone gauge),
where the second factor corresponds to a global R-symmetry from
the six-dimensional point of view. For Type IIA the resulting
six-dimensional theory has ${\cal N}=(1,1)$ supersymmetry,
and for Type IIB it has ${\cal N}=(2,0)$ supersymmetry.
The relevant properties of the different sectors are shown in
table~2. 
\begin{table}[htb]
\begin{tabular}{|r|l|l|l|}
\hline
 sector & $\alpha_n^i$ & $\psi_r^i$ & $SO(4)_S\times SO(4)_R$ \\
\hline
 &&&\\[-10pt]
 NS;U & $n\in\bbbz$ & $r\in\bbbz + 1/2$ & $({\bf 1},{\bf 1})$ \\[5pt]
 R;U & $n\in\bbbz$ & $r\in\bbbz$ & 
    $({\bf 2}\oplus{\bf 2}',{\bf 2}\oplus{\bf 2}')$ \\[5pt]
 NS;T & $n\in\left\{
 \begin{array}{ll}
  \bbbz & i=1,\ldots,4 \\
  \bbbz + 1/2 & i=5,\ldots,8
 \end{array}\right.$
 & $r\in\left\{
 \begin{array}{ll}
  \bbbz + 1/2 & i=1,\ldots,4 \\
  \bbbz & i=5,\ldots,8
 \end{array}\right.$
 & $({\bf 1},{\bf 2}\oplus{\bf 2}')$ \\[20pt]
 R;T & $n\in\left\{
 \begin{array}{ll}
  \bbbz & i=1,\ldots,4 \\
  \bbbz + 1/2 & i=5,\ldots,8
 \end{array}\right.$
 & $r\in\left\{
 \begin{array}{ll}
  \bbbz & i=1,\ldots,4 \\
  \bbbz + 1/2 & i=5,\ldots,8
 \end{array}\right.$
 & $({\bf 2}\oplus{\bf 2}',{\bf 1})$ \\[10pt]
\hline
\end{tabular}
\caption{\small
Oscillator modings in light-cone gauge and ground state charges of the
different sectors in the orbifold $\bbbr^4/\bbbz_2$. The ground state
is tachyonic in  the NS;U sector, and massless in all the other
sectors. Its charges are determined by quantizing the fermionic zero
modes.} 
\end{table}

Modular invariance requires that we project all sectors onto states
which are even under $g$, 
\be
 P_{orbifold} = {1\over 2}(1+ g) \;.
\label{orb-projection}
\ee
In principle, modular invariance should also determine the correct 
GSO-projections in the twisted sectors, once that of the untwisted R-R
sector has been fixed, {\it i.e.} once we have decided whether to
start with Type IIA or Type IIB string theory. However, due to the
sign ambiguity in the action of the GSO operators on the ground states
of these sectors, it is actually difficult to determine the consistent
projections. The most general GSO-projections are given
as\footnote{Only the relative phase of $(-1)^f$ and
$(-1)^{\widetilde{f}}$ is relevant 
for our purposes}:
\be
 P_{GSO} = \left\{
 \begin{array}{ll}
  {1\over 4}(1+(-1)^f)(1+(-1)^{\widetilde{f}}) & \mbox{NS-NS;U} \\
  {1\over 4}(1+(-1)^f)(1+\epsilon (-1)^{\widetilde{f}}) 
          & \mbox{R-R;U}\\ 
  {1\over 4}(1+(-1)^f)(1+\eta (-1)^{\widetilde{f}}) 
          & \mbox{NS-NS;T} \\
  {1\over 4}(1+(-1)^f)(1+\delta (-1)^{\widetilde{f}}) 
          & \mbox{R-R;T},
 \end{array}\right.
\label{GSO-projection}
\ee
where $\epsilon, \eta, \delta = \pm 1$. The phase $\epsilon$
corresponds to whether the theory is Type IIA ($\epsilon =-1$) or Type
IIB  ($\epsilon= +1$), and the phases $\eta,\delta$ correspond to the
aforementioned ambiguity.

To fix the ambiguity in the choice for $\eta$ and $\delta$, let us now
consider the D-brane spectrum of the theory. From the projections in
the untwisted sectors it follows that the physical bulk D-branes have
$r$ even and $s$ even in Type IIA, and $r$ odd and $s$ even in Type
IIB. In order to satisfy the above fractionation condition, there
must exist fractional D-branes for the same values of $r$ and $s$. 
In particular,
this means that the corresponding boundary states must be physical in
both the twisted NS-NS and the twisted R-R sectors. It follows from
the results of Table~1 that this requires $\eta=+1$ and 
$\delta=\eta\epsilon=\epsilon$; the D-brane spectrum can then be
summarised by 
\begin{list}{(\roman{enumi})}{\usecounter{enumi}}
\item {\bf IIA on $\bbbr^4/\bbbz_2$}: Fractional and bulk D-branes exist
for $r$ and $s$ both even. \\
Truncated D-branes exist for $r$ even and  $s$ odd,
{\em e.g.} the non-BPS D-string \cite{Sen6,BG3}.
\item {\bf IIB on $\bbbr^4/\bbbz_2$}: Fractional and bulk D-branes exist
for $r$ odd and $s$ even. \\
Truncated D-branes exist for both $r$ and $s$ odd. 
\end{list}

This choice of $\eta$ and $\delta$ also leads to a supersymmetric
spectrum in the twisted sectors. Indeed, the surviving components of
the ground state in the twisted R-R sector transform as 
\be
 ({\bf 2},{\bf 1})\otimes ({\bf 2}',{\bf 1}) = ({\bf 4},{\bf 1})
\label{vector}
\ee
in the Type IIA orbifold, and as
\be
 ({\bf 2},{\bf 1})\otimes ({\bf 2},{\bf 1}) = ({\bf 3},{\bf 1})\oplus
  ({\bf 1},{\bf 1})
\label{tensor}
\ee
in the Type IIB orbifold. The former corresponds to the vector component
of an ${\cal N}=(1,1)$ vector multiplet, and the latter to the rank two
antisymmetric tensor component and one of the scalar components of 
an ${\cal N}=(2,0)$ tensor multiplet. As these are precisely the
unbroken supersymmetries associated with the respective theories,
the above choice for the GSO-projection in the twisted sectors is
consistent with supersymmetry as well. 

It may also be worth mentioning that only this choice for the 
GSO-projection in the twisted sectors leads to an orbifold that can be
blown up to a smooth ALE space (K3 in the fully compact case),
since the latter is manifestly supersymmetric.

\paragraph{4. Non-supersymmetric example.} 

Now consider the case of Type 0 strings on 
$\bbbr^{1,3}\times\bbbr^6/\bbbz_2$, where the generator $g$ of 
$\bbbz_2$ reflects the six coordinates $x^3,\ldots,x^8$.\footnote{This
case is of particular interest since it is directly related to recent
work of Klebanov, Nekrasov and Shatashvili \cite{KNS}, where this
orbifold of Type 0B was discussed.} Note that $g^2=(-1)^F$, but since
Type 0 is purely bosonic this is equivalent to the identity. In Type
II strings $g$ generates a $\bbbz_4$ group, so the above theory is
identical to Type II on $\bbbr^{1,3}\times\bbbr^6/\bbbz_4$. The
ten-dimensional Lorentz group is broken to $SO(1,3)\times SO(6)$ 
($SO(2)\times SO(6)$ in light-cone gauge), and the theory is not
supersymmetric (and in fact completely fermion-free).
The different sectors of the theory are described in table~3.
\begin{table}[htb]
\begin{tabular}{|r|l|l|l|}
\hline
 sector & $\alpha_n^i$ & $\psi_r^i$ & $SO(2)_S\times SO(6)_R$ \\
\hline
 &&&\\[-10pt]
 NS;U & $n\in\bbbz$ & $r\in\bbbz + 1/2$ & ${\bf 1}_0$ \\[5pt]
 R;U & $n\in\bbbz$ & $r\in\bbbz$ & 
    $({\bf 4}\oplus\overline{\bf 4})_{1\over 2}\oplus 
     ({\bf 4}\oplus\overline{\bf 4})_{-{1\over 2}}$ \\[5pt]
 NS;T & $n\in\left\{
 \begin{array}{ll}
  \bbbz & i=1,2 \\
  \bbbz + 1/2 & i=3,\ldots,8
 \end{array}\right.$
 & $r\in\left\{
 \begin{array}{ll}
  \bbbz + 1/2 & i=1,2 \\
  \bbbz & i=3,\ldots,8
 \end{array}\right.$
 & ${\bf 4}_0 \oplus \overline{\bf 4}_0$ \\[20pt]
 R;T & $n\in\left\{
 \begin{array}{ll}
  \bbbz & i=1,2 \\
  \bbbz + 1/2 & i=3,\ldots,8
 \end{array}\right.$
 & $r\in\left\{
 \begin{array}{ll}
  \bbbz & i=1,2 \\
  \bbbz + 1/2 & i=3,\ldots,8
 \end{array}\right.$
 & ${\bf 1}_{1\over 2} \oplus {\bf 1}_{-{1\over 2}}$ \\[10pt]
\hline
\end{tabular}
\caption{\small
Oscillator modings and ground state charges of the different sectors
in the orbifold $\bbbr^6/\bbbz_2$.
The ground state is tachyonic in the NS;U sector, massive in the NS;T 
sector, and massless in the R;U and R;T sectors.}
\end{table}

Since we are using the Type 0 picture, the relevant GSO operator is
actually the combination $(-1)^{f+\widetilde{f}}$. But since the
action of $(-1)^f$ is trivial on all boundary states, we can still
refer to table~1 (with $n=6$ in this case) for the transformation
properties of the boundary states.\footnote{In fact, both $(-1)^f$ 
and $(-1)^{\widetilde{f}}$ change the spin structure of the boundary
states, and the Type 0 GSO operator $(-1)^{f+\widetilde{f}}$ therefore 
preserves the spin structure. Thus it is not necessary to consider
a linear combination of boundary states, and this gives rise to the
famous doubling of the D-brane spectrum of Type 0 relative to 
Type II.} The most general GSO-projections are now given by  
\be
 P_{GSO} = \left\{
 \begin{array}{ll}
  {1\over 2}(1+(-1)^{f+\widetilde{f}}) & \mbox{NS-NS} \\
  {1\over 2}(1+\epsilon(-1)^{f+\widetilde{f}}) & \mbox{R-R} \\
  {1\over 2}(1+\eta(-1)^{f+\widetilde{f}}) & \mbox{NS-NS;T} \\
  {1\over 2}(1+\delta(-1)^{f+\widetilde{f}}) & \mbox{R-R;T}\;,
 \end{array}\right.
\ee
where $\epsilon$ again determines whether the theory is Type A or B.

Referring again to table~1, it now follows that in order to satisfy
the D-brane fractionation condition, we must have $\eta =-1$,
and $\delta = \eta\epsilon = -\epsilon$.
For this choice (and only for this choice) we obtain a spectrum of 
D-branes that is consistent with the above fractionation condition; 
this D-brane spectrum can be summarised by 
\begin{list}{(\roman{enumi})}{\usecounter{enumi}}
\item {\bf 0A on $\bbbr^6/\bbbz_2$}: Fractional and bulk D-branes
exist for $r$ and $s$ both even. \\
Truncated D-branes exist for $r$ even and $s$ odd.
\item {\bf 0B on $\bbbr^6/\bbbz_2$}: Fractional and bulk D-branes
exist for $r$ odd and $s$ even. \\
Truncated D-branes exist for both $r$ and $s$ odd. 
\end{list}

The above solutions for $\eta$ and $\delta$ are somewhat
counterintuitive, in that the GSO-projection in the twisted sectors is
opposite to that in the untwisted sectors. In particular, the
projection in the twisted R-R sector is {\em chiral} in Type 0A and
{\em non-chiral} in Type 0B, which is opposite to the convention in
the untwisted R-R sector. The surviving massless fields in the twisted
R-R sector therefore transform as 
\be
 ({\bf 1}_{1\over 2}\otimes {\bf 1}_{{1\over 2}}) \oplus
 ({\bf 1}_{-{1\over 2}}\otimes {\bf 1}_{-{1\over 2}}) =
 {\bf 1}_1 \oplus {\bf 1}_{-1}
\ee
in the Type 0A orbifold, and as
\be
 ({\bf 1}_{1\over 2}\otimes {\bf 1}_{-{1\over 2}}) \oplus
 ({\bf 1}_{-{1\over 2}}\otimes {\bf 1}_{{1\over 2}}) =
 2\times {\bf 1}_0 
\ee
in the Type 0B orbifold. The former corresponds to the two helicities 
of a massless vector in four dimensions, and the latter to two
massless scalars.
\footnote{For the case of the Type 0B orbifold, this also agrees with
the convention that was considered in \cite{BK}.}

This is the opposite of what was claimed in \cite{KNS}, namely that
the twisted sector of the Type 0B orbifold contains a massless
vector. In effect, the authors of \cite{KNS} chose $\eta=\delta=+1$,
and therefore a chiral GSO-projection in the twisted R-R sector. It is
clear from table~1 that in this case there are no physical fractional
branes, and therefore that the fractionation condition is violated.

\paragraph{5. Conclusions.}

The condition of modular invariance, which underlies all consistent
closed string theories, is sometimes clouded by phase ambiguities.
This is especially true for the GSO-projection in sectors containing
fermionic zero modes. In this note we have explained that there
exists a non-perturbative consistency condition for orbifold theories,
related to the spectrum of D-branes, that determines the
GSO-projection in the twisted sectors uniquely. For the supersymmetric
cases, this condition reproduces the conventions that follow from
supersymmetry, but it also applies to situations where supersymmetry
is broken. As an example of the latter case, we considered Type 0
strings on $\bbbr^{1,3}\times\bbbr^6/\bbbz_2$.  In this case,
requiring the spectrum of fractional D-branes to match up with that of
the bulk D-branes fixes the GSO-projection in the twisted sectors, and
thus the spectrum of the theory.

\paragraph{Acknowledgments.}

We thank Igor Klebanov for useful correspondences. M.R.G. thanks Peter
Goddard for a helpful conversation. O.B. thanks Zurab Kakushadze.
O.B. is supported in part by the DOE under grant no. DE-FG03-92-ER
40701. M.R.G. is grateful to the Royal Society for a University
Research Fellowship.


\begin{thebibliography}{[20]}

\bibitem{DHVW} L. Dixon, J.A. Harvey, C. Vafa, E. Witten, {\it Strings
on orbifolds I \& II}, Nucl. Phys. {\bf B261}, 678 (1985) and
Nucl. Phys. {\bf B274}, 285 (1986).

\bibitem{Vafa} C. Vafa, {\it Modular invariance and discrete torsion
on orbifolds}, Nucl. Phys. {\bf B273}, 592 (1986).

\bibitem{PolCai} J. Polchinski, Y. Cai, {\it Consistency of open
superstring theories}, Nucl. Phys. {\bf B296}, 91 (1988).

\bibitem{CLNY} C.G. Callan, C. Lovelace, C.R. Nappi, S.A. Yost,
{\it Loop corrections to superstring equations of motion},
Nucl. Phys. {\bf B308}, 221 (1988).

\bibitem{Cardy} J.L. Cardy, {\it Boundary conditions, fusion rules,
and the Verlinde formula}, Nucl. Phys. {\bf B324}, 581 (1989).

\bibitem{Lew} D. Lewellen, {\it Sewing constraints for conformal field
theories on surfaces with boundaries}, Nucl. Phys. {\bf B372}, 654
(1992). \\
J.L. Cardy, D. Lewellen, {\it Bulk and boundary operators in conformal
field theory}, Phys. Lett. {\bf B259}, 274 (1991). 

\bibitem{BG1} O. Bergman, M.R. Gaberdiel, {\it A non-supersymmetric
open string theory and S-duality}, Nucl. Phys.~{\bf B499}, 183 (1997); 
{\sf hep-th/9701137}. 

\bibitem{GS} M.R. Gaberdiel, B. Stefa\'nski, {\it Dirichlet Branes on
Orbifolds}, {\sf hep-th/9910109}.

\bibitem{DM} M.R. Douglas, G. Moore, {\it D-branes, quivers, and
ALE instantons}, {\sf hep-th/9603167}.

\bibitem{Sen2} A. Sen, {\it Stable non-BPS bound states of BPS
D-branes}, JHEP {\bf 9808}, 010 (1998); {\sf hep-th/9805019}.

\bibitem{BG2} O. Bergman, M.R. Gaberdiel, {\it Stable non-BPS
D-particles}, Phys. Lett.~{\bf B441}, 133 (1998);
{\sf hep-th/9806155}.

\bibitem{Sen6} A. Sen, {\it BPS D-branes on non-supersymmetric
cycles}, JHEP {\bf 9812}, 021 (1998); {\sf hep-th/9812031}.

\bibitem{BG3} O. Bergman, M.R. Gaberdiel, {\it  Non-BPS states in
Heterotic -- Type IIA duality}, JHEP {\bf 9903}, 013 (1999); 
{\sf hep-th/9901014}. 

\bibitem{DG} D-E. Diaconescu, J. Gomis, {\it Fractional branes and
boundary states in orbifold theories}, {\sf hep-th/9906242}. 

\bibitem{GabSen} M.R. Gaberdiel, A. Sen, {\it Non-supersymmetric
D-Brane configurations with Bose-Fermi degenerate open string
spectrum}, JHEP {\bf 9911}, 008 (1999); {\sf hep-th/9908060}.

\bibitem{KNS} I.R. Klebanov, N.A. Nekrasov, S.L. Shatashvili, {\it An
orbifold of Type 0B strings and non-supersymmetric gauge theories},
{\sf hep-th/9909109}.

\bibitem{BK} R. Blumenhagen, A. Kumar, {\it A Note on Orientifolds and
Dualities of Type 0B String Theory}, Phys. Lett. {\bf B464}, 46
(1999); {\sf hep-th/9906234}.


\end{thebibliography}
\end{document}